%% file: main.tex
\pdfoutput=1
\documentclass[11pt]{article}

\PassOptionsToPackage{hyphens}{url}

\usepackage[preprint]{acl}

\usepackage{times}
\usepackage{latexsym}
\usepackage[T1]{fontenc}
\usepackage[utf8]{inputenc}
\usepackage{microtype}
\usepackage{inconsolata}
\usepackage{graphicx}
\usepackage{amsmath}     

\usepackage{booktabs}    
\usepackage{colortbl}    
\usepackage{xcolor}
\usepackage{verbatim}    
\usepackage{xspace}      
\usepackage{algorithm}
\usepackage{algpseudocode}
\usepackage{listings}
\definecolor{ourrow}{HTML}{E8F0FE}  

\newcommand{\system}{PatchHolmes\xspace}
\newcommand{\tool}[1]{\texttt{#1}}
\newcommand{\fcall}[2]{\tool{#1}\allowbreak\texttt{(#2)}}

\title{\system: Agentic Patch Retrieval via Listwise Selection\thanks{Accepted at AACL-IJCNLP 2026. This is the authors' preprint version.}}

\author{Guanqun Yang\thanks{Equal contribution.} \quad
  Yingming Zhou\footnotemark[2] \quad
  Jiangrui Zheng\footnotemark[2] \quad
  Shudong Hao \quad
  Xueqing Liu \\
  Stevens Institute of Technology, Hoboken, NJ, USA \\
  \texttt{guanqun.yang@outlook.com}, \texttt{\{yzhou136, jzheng36, shao14, xliu127\}@stevens.edu}}

\begin{document}
\maketitle

\begin{abstract}
Patch retrieval, the task of finding the commit that fixes a known vulnerability, is the foundation of vulnerability management workflows, yet 60\% to 63\% of CVEs in the major advisory databases lack a patch link.
We present \system, a two-phase patch retrieval system that pairs a hybrid first-stage retriever with an agentic second-stage inspection loop.
Unlike pointwise prior work that scores each candidate independently, the Phase 2 agent reads the top-100 listwise: it sees the full candidate list at once and selectively reads 3 to 10 commits through four budgeted tools before submitting a single best commit.
On GitHubAD, \system\ beats the pointwise binary classifier Favia by 25.34\% Recall@1 and the retrieve-and-CoT baseline IRCoT by 31.40\%, at one agent conversation per CVE versus Favia's ten; with the candidate set held identical, the agent adds 27.32\% Recall@1 over taking the retriever's top candidate, and the same agent, transferred unchanged to \mbox{PatchFinder\_top10}, lifts Recall@1 from PatchFinder's own top-1 pick (24.28\%) to 39.86\%.
Swapping the LLM backbone within the Qwen family changes Recall@1 by under 1\%, and a second model family (gpt-oss) stays far above the no-agent floor, so the gain comes from the listwise agent loop; the entire system runs on a frozen open-weight model over a local Git repository, without fine-tuning or external search APIs.
\end{abstract}

\input{sections/001-intro.tex}
\input{sections/002-related.tex}
\input{sections/003-method.tex}
\input{sections/004-experiments.tex}
\input{sections/005-conclusion.tex}

\clearpage
\input{sections/006-limitations.tex}

\bibliography{zotero}

\appendix

\input{sections/900-appendix.tex}

\end{document}

%% file: sections/001-intro.tex
\section{Introduction}
\label{sec:intro}

\begin{figure}[!t]
  \centering
  \includegraphics[width=\columnwidth]{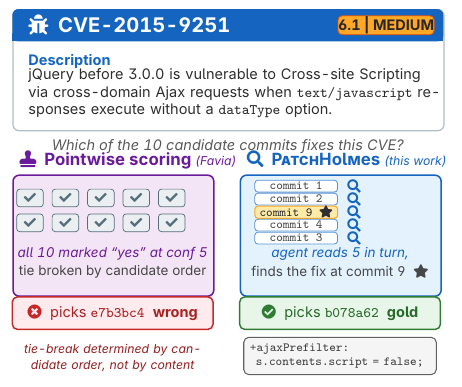}
  \caption{\textbf{Listwise Selection (\system) Recovers a Pointwise Failure (Favia) on \texttt{CVE-2015-9251} (jQuery XSS via Cross-Domain Ajax).} On the same Phase 1 top-10 shortlist (gold at rank 9), Favia gives all 10 candidates identical confidence and picks the rank-1 refactor; \system's agent reads five commits in Phase 1 rank order (1, 2, \textbf{9}, 4, 3) and identifies the one that adds the line blocking automatic execution of \texttt{text/javascript} responses.}
  \label{fig:case-study}
\end{figure}

Vulnerability management rests on a basic mapping: every disclosed software vulnerability needs to be paired with the commit that fixes it in the upstream repository~\citep{Dunlap2024VFCFinderPairingSecurity, Tan2021LocatingSecurityPatchesa}.
The task, called \emph{patch retrieval}~\citep{Li2024PatchFinderTwoPhaseApproach, Zheng2026SPFinderImprovingContext, Storhaug2026FaviaForensicAgent}, asks for the commit identifier that addresses a given CVE in a given repository.
Downstream consumers include security advisories, vulnerability-assessment scores (CVSS), affected-version-range trackers, and SBOM-driven supply-chain scanners~\citep{ODonoghue2024ImpactsSoftwareBill}.
However, the supply of patch links does not keep up with demand.
Independent studies report that 60\% to 63\% of CVEs in the GitHub Advisory Database and the NVD are missing their patch links~\citep{Dunlap2024VFCFinderPairingSecurity, Zheng2026SPFinderImprovingContext}, and auxiliary-information-based matching can recover only 12\% to 53\% of disclosed OSS vulnerabilities even with manual assistance~\citep{Tan2021LocatingSecurityPatchesa}.
Producing the mapping manually does not scale, and the NVD itself faces a maintainer backlog that delays metadata curation~\citep{Liu2024CanHighlightingHelp}.
The missing-link problem can persist for years; the patch for CVE-2013-1814 in Apache Rave was not surfaced until a decade after disclosure.\footnote{\url{https://nvd.nist.gov/vuln/detail/CVE-2013-1814}}
A patch retrieval system intended to scale with this backlog must run on a frozen LLM over a local Git repository, without per-repository fine-tuning and without a paid external search API, because the cost of either option grows linearly with the CVE volume.

Three properties of the setting make patch retrieval inherently hard.
First, the corpus is large: 49\% of CVEs are filed against repositories with more than 5,000 commits~\citep{Zheng2026SPFinderImprovingContext}.
Second, the diffs are long: in our 8,401-CVE GitHubAD corpus, the average commit diff runs about 15,000 tokens, well past the 512-token window of the pre-LLM encoders such as BERT~\citep{Devlin2018} that prior methods rely on.
Third, the CVE description shares relatively little vocabulary with the commit message that fixes it~\citep{Tan2021LocatingSecurityPatchesa}: a CVE typically names the security symptom (``buffer overflow''), while the patch commit names the underlying memory-access bug (``out-of-bounds read''), so a lexical retriever cannot match them on shared tokens.
A first-stage retriever can surface the right commit somewhere in its top-100, but the gold patch can sit at any rank within that 100. The bottleneck is therefore the second-stage selector that decides which of those 100 is the actual fix.

A line of \emph{traditional} patch retrieval~\citep{Tan2021LocatingSecurityPatchesa, Dunlap2024VFCFinderPairingSecurity, Li2024PatchFinderTwoPhaseApproach, Zheng2026SPFinderImprovingContext} has made substantial progress on this selector, but has two fundamental limitations.
First, it is \emph{open-loop}: a static feature stack scores every candidate once and emits the final ranking, with no way to refine the query when the top is wrong, inspect a candidate in detail, or back out and try again.
Second, its encoders are \emph{pre-LLM}: the 512-token BERT-class context window cannot fit the 15,000-token average diff, so most of the diff is truncated before scoring even begins.
An agentic second stage closes both gaps in principle: a multi-turn agent can refine its query based on the observation returned by the previous tool call, and tool-mediated diff rendering lets it read long commits in budgeted chunks rather than rely on encoder-level truncation.

Two recent agentic lines come close, but each has a limitation.
\emph{General-domain} closed-loop retrievers~\citep{Trivedi2023InterleavingRetrievalChainofThought, Jiang2023ActiveRetrievalAugmented, Asai2023SelfRAGLearningRetrieve, Jin2025SearchR1TrainingLLMs} alternate retrieval and reasoning over text without any task-specific tooling. Because they are tuned for open-web multi-hop QA where every hop shares vocabulary, they transfer less directly to the closed-world setting here.
Off-the-shelf IRCoT~\citep{Trivedi2023InterleavingRetrievalChainofThought} is the only such retriever that runs without fine-tuning and without a paid external search API, the same constraint our setting imposes, which makes it the appropriate general-domain baseline; it reaches only 28.55\% Recall@1 in our setup.
On the \emph{custom-built} side, one recent system, Favia~\citep{Storhaug2026FaviaForensicAgent}, replaces the fine-tuned scorer of the traditional line with an LLM that emits a yes/no plus a confidence for each of the 10 pre-filtered candidates per CVE.
Favia's design is structurally pointwise: each of the 10 calls is independent of the other nine, so when multiple candidates plausibly receive yes the derived rank-1 pick has no joint signal to break the tie.
Consider \texttt{CVE-2015-9251}, a jQuery cross-site scripting via cross-domain Ajax (Figure~\ref{fig:case-study}): Favia's 10 pointwise calls mark every candidate as ``yes'' at confidence 5, and the derived rank-1 pick (\texttt{e7b3bc4}, an unrelated refactor of the cross-domain code path) is decided by candidate order rather than by content, missing the actual fix.

We address these limitations with \system, which performs patch retrieval with \emph{listwise selection}.
A hybrid Phase 1 retriever produces a top-100 candidate set; an agent then reads the top-100 through four budgeted tools and submits a single best commit per CVE, comparing candidates against each other rather than scoring them independently.
The multi-turn loop refines the agent's next action on each tool observation, replacing the open-loop scorer. Tool-mediated chunked diff rendering lets the agent read past the 512-token encoder window, and \tool{list\_candidates} lets the agent see all 100 candidates at once before drilling into any single commit.
Returning to \texttt{CVE-2015-9251}: \system's agent reads five commits in P1-rank order (1, 2, 9, 4, 3) through \tool{read\_commit} and identifies \texttt{b078a62} as the fix, because it alone adds the \texttt{ajaxPrefilter} that blocks the automatic execution of \texttt{text/javascript} responses for cross-domain Ajax requests.
One agent conversation per CVE replaces Favia's ten.

\begin{figure*}[t]
  \centering
  \includegraphics[width=\textwidth]{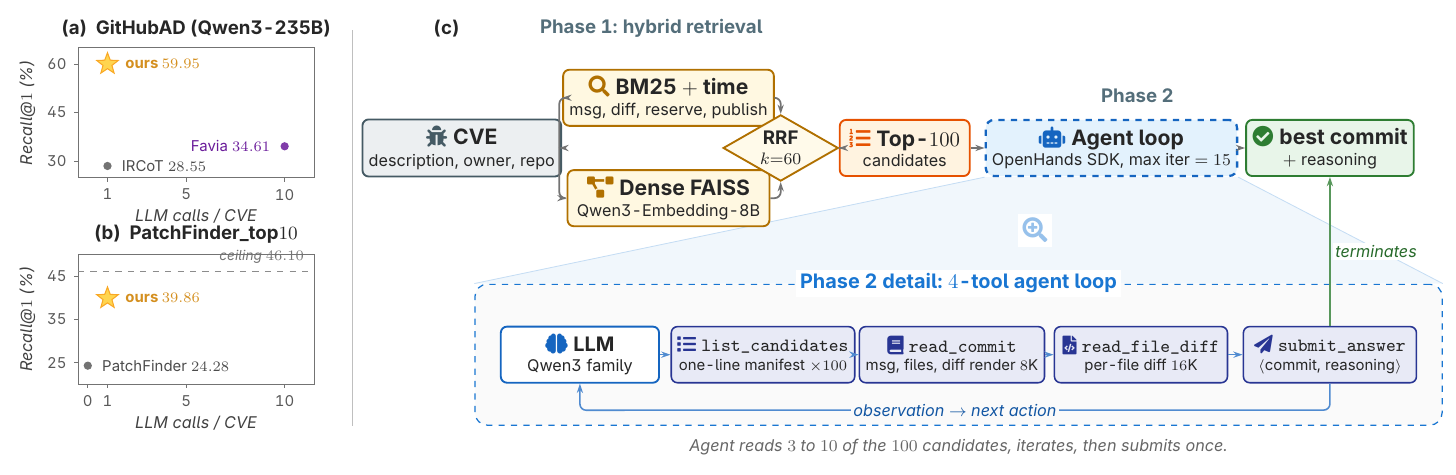}
  \caption{\textbf{\system\ Traces a CVE to Its Fix Commit in Two Phases.} Panels~(a, b) plot accuracy against LLM-call cost; panel~(c) is the system diagram. On GitHubAD~\citep{Zheng2026SPFinderImprovingContext}, \system\ (gold star) Pareto-dominates Favia~\citep{Storhaug2026FaviaForensicAgent} and IRCoT~\citep{Trivedi2023InterleavingRetrievalChainofThought} at Recall@1 of 59.95\% with one agent conversation per CVE, against Favia's 34.61\% at ten. On the \mbox{PatchFinder\_top10} benchmark~\citep{Storhaug2026FaviaForensicAgent}, \system\ reaches Recall@1 of 39.86\% against PatchFinder~\citep{Li2024PatchFinderTwoPhaseApproach} at 24.28\%, approaching the 46.10\% recoverable ceiling. Phase 1 fuses BM25-with-time-decay and a Qwen3-Embedding-8B dense retriever via Reciprocal Rank Fusion into a top-100 candidate set; Phase 2 is an agent that drives four tools (\tool{list\_candidates}, \tool{read\_commit}, \tool{read\_file\_diff}, \tool{submit\_answer}) over up to 15 iterations and submits a single best commit per CVE, typically after reading 3 to 10 candidates.}
  \label{fig:system}
\end{figure*}

\paragraph{Contribution.}
We present \system, a two-phase patch retrieval system with \emph{listwise selection}: a hybrid first-stage retriever paired with an agentic second-stage inspection loop driving four commit-reading tools, running on a frozen open-weight LLM over a local Git repository with no fine-tuning and no external search API.
On GitHubAD~\citep{Zheng2026SPFinderImprovingContext}, \system\ beats Favia~\citep{Storhaug2026FaviaForensicAgent} by 25.34\% Recall@1 and the canonical retrieve-and-CoT baseline~\citep{Trivedi2023InterleavingRetrievalChainofThought} by 31.40\%, at one agent conversation per CVE (\S\ref{sec:main}).
The benefit transfers to the \mbox{PatchFinder\_top10} candidate pool: the same agent, applied without changes, lifts Recall@1 from PatchFinder's trivial rank-1 baseline of 24.28\% to 39.86\% with Phase 1 skipped, recovering 71.40\% of the gap to the recoverable ceiling (\S\ref{sec:cross-corpus}).
On identical candidates, the agent alone adds 27.32\% Recall@1 over the retriever's top pick, across two model families (\S\ref{sec:analysis}), at about 97K input tokens per CVE (\S\ref{sec:cost}).
We open-source \system\ at \url{https://github.com/Aizhouym/PatchHolmes}.

%% file: sections/002-related.tex
\section{Related Work}
\label{sec:related}

\paragraph{Patch Retrieval.}
PatchScout~\citep{Tan2021LocatingSecurityPatchesa} established the feature-based ranking template, scoring commits by hand-engineered vulnerability-commit correlations over file paths, types, and identifiers.
The two-phase template of patch retrieval was established by PatchFinder~\citep{Li2024PatchFinderTwoPhaseApproach}: a lexical-and-neural first-stage retriever feeds a fine-tuned cross-encoder over the per-CVE top-100.
SPFinder~\citep{Zheng2026SPFinderImprovingContext} keeps the two-phase shape and addresses the long-diff problem with a hierarchical embedding and a feature-based gradient-boosted reranker.
Favia~\citep{Storhaug2026FaviaForensicAgent} replaces the fine-tuned scorer with an LLM that emits a pointwise yes/no and confidence per candidate.
\system\ targets the same setting, but replaces SPFinder's open-loop feature stack and Favia's independent pointwise calls with a single agentic loop that reads the candidate manifest jointly and drills into individual commits selectively (\S\ref{sec:main}).

\paragraph{Reasoning-Intensive Retrieval.}
Open-domain retrieval typically admits many relevant documents; patch retrieval has exactly one correct commit out of up to 1.4M, with little vocabulary overlap between the CVE description and the commit that fixes it~\citep{Storhaug2026FaviaForensicAgent}.
This is the regime that the BRIGHT benchmark~\citep{Su2025BRIGHTRealisticChallenging} constructs, on which classical retrievers collapse and reasoning-augmented retrievers gain the most.
\citet{Chen2025BrowseCompPlusMoreFaira} push the diagnosis further on a deep-research benchmark: holding the LLM agent fixed, swapping the retriever changes end-to-end accuracy more than swapping the agent does.
\system's empirical finding that the agent loop adds 27.32\% Recall@1 on fixed candidates while a backbone swap moves it by under 1\% within a family (\S\ref{sec:analysis}) is consistent with this picture.

\paragraph{Coding and Retrieval Agents.}
\system's Phase 2 instantiates the ReAct paradigm~\citep{Yao2023ReActSynergizingReasoning} of interleaving reasoning and tool use over multiple turns.
On natural-language text, this loop has been refined by alternating retrieval with chain-of-thought~\citep{Trivedi2023InterleavingRetrievalChainofThought}, re-retrieving on the model's own uncertainty~\citep{Jiang2023ActiveRetrievalAugmented}, training a critic that decides when to retrieve~\citep{Asai2023SelfRAGLearningRetrieve}, and learning the loop end-to-end with reinforcement learning over a search engine~\citep{Jin2025SearchR1TrainingLLMs, chen2025researchlearningreasonsearch}.
Closer to our setting, the same loop has been extended to source-code repositories: a deep-research agent over commit history and code-symbol search for Linux-kernel crashes~\citep{Singh2026CodeResearcherDeep}, a repository compiled into a knowledge graph and searched via a query language~\citep{Shah2025RANGERRepositoryLevelAgenta}, and a repository-navigation agent trained with reinforcement learning~\citep{Ma2025ToolintegratedReinforcementLearninga}.
Among agentic systems, only Favia targets patch retrieval directly~\citep{Storhaug2026FaviaForensicAgent}; we benchmark \system\ against it head-to-head in \S\ref{sec:main}.

%% file: sections/003-method.tex
\section{Method}
\label{sec:method}

\system\ traces a CVE to its fix commit in two phases (Figure~\ref{fig:system}c).
The input is a CVE description $q$ and a target repository $R$ with commits $\{c_1, \ldots, c_N\}$, typically in the thousands to millions of commits. Phase 1 is recall-oriented: it retrieves a 100-candidate shortlist $C \subset R$ from $q$ alone. Phase 2 is precision-oriented: it inspects $C$ with an LLM agent that selects a single $\hat{c} \in C$ as the predicted fix.
The two phases share no state beyond the candidate identifiers: Phase 1 runs once per CVE and writes a JSONL record, and Phase 2 reads from it.

\subsection{Phase 1: Hybrid Retrieval}
\label{sec:phase1}

Phase 1 runs two retrieval paths in parallel and fuses their rankings with Reciprocal Rank Fusion.

\paragraph{Path 1: BM25 with Time Decay.}
The lexical path scores every commit on four signals and combines them by fixed weight:
\begin{align*}
s_{\text{lex}}(c) \,=\,& \,0.35\,b_{\text{msg}}(c) + 0.15\,b_{\text{diff}}(c) \\
                       & + 0.30\,t_{\text{reserve}}(c) + 0.20\,t_{\text{publish}}(c).
\end{align*}
$b_{\text{msg}}$ and $b_{\text{diff}}$ are BM25 scores over the commit message and the diff body respectively, each min-max normalized against the top 10,000 hits in its field.
$t_{\text{reserve}}$ and $t_{\text{publish}}$ are time-rank scores that compare a commit's datetime against the CVE's reserve date and public-disclosure date.
The time-rank score for a commit at local rank $r$ relative to the CVE at local rank $c$ (both indexed within the top-10,000 BM25 candidates sorted by datetime) is $1/(1 + 2|r-c|)$; the score peaks at the commit closest in time to the CVE and decays symmetrically (from $1$ at $|r-c|=0$ to $1/3$ at $|r-c|=1$ to $1/21$ at $|r-c|=10$).
The rank-difference form lets us combine the time signal directly with reciprocal-rank-fused lexical scores without re-calibrating absolute time deltas.
The CVE-reserve-to-commit time difference is a strong relevance signal introduced by SPFinder~\citep{Zheng2026SPFinderImprovingContext}, which also provides the specific four weights $(0.35, 0.15, 0.30, 0.20)$ via grid search over that signal.
The largest coefficient (0.35) goes to the commit-message BM25 because commit messages share more vocabulary with the CVE description than the raw diff does; the diff-body BM25 gets the smallest (0.15) because its tokens are mostly code that does not overlap with the natural-language CVE text.

\paragraph{Path 2: Dense Retrieval.}
The dense path embeds the CVE description and every commit into a shared 4096-dimensional space with Qwen3-Embedding-8B~\citep{Zhang2025Qwen3EmbeddingAdvancing} and ranks by cosine similarity.
We chose Qwen3-Embedding-8B over the RTEB-leading\footnote{\url{https://huggingface.co/spaces/embedding-benchmark/RTEB}} Octen-Embedding-8B (a LoRA fine-tuned model of the same backbone) on the basis of a pilot ablation: swapping Qwen3-Embedding-8B for Octen drops Phase 1 dense-leg Recall@1 from 43.39\% to 12.73\% on our task (Appendix~\ref{sec:dense-ablation-appendix}), so the RTEB-leading fine-tuning does not transfer to patch retrieval.
For the diff, we sort lines into four buckets by information density (file paths $>$ hunk headers $>$ $\pm$ lines $>$ context lines) and fill a 6,000-character budget bucket by bucket from highest priority down, keeping whole lines only so the encoder never sees a half-line.
In pilot experiments, we found that 6,000 characters is the best budget: relaxing it past the encoder's effective window degrades retrieval quality faster than the extra context recovers it.
Embeddings are precomputed offline per repository and cached as a single matrix; per-CVE retrieval is then one matrix multiplication against that cache rather than a per-query index build.
This offline-indexing pattern is standard industry practice in production code-search systems such as Sourcegraph's Zoekt\footnote{\url{https://github.com/sourcegraph/zoekt}} and GitHub's Blackbird.\footnote{\url{https://github.blog/2021-12-08-improving-github-code-search/}}

\paragraph{RRF Fusion and Truncation.}
The two full rankings are fused with Reciprocal Rank Fusion at $k{=}60$:
\[
s_{\text{RRF}}(c) \,=\, \frac{1}{60 + r_{\text{lex}}(c)} + \frac{1}{60 + r_{\text{dense}}(c)}.
\]
The top-1000 by fused score is written to disk; Phase 2 reads the first 100 of those.
We set the cut at 100 because it lets the agent see deep enough into the ranking to recover gold commits the open-loop scorer ranks late, while keeping the conversation prompt short enough to stay under one LLM call's worth of commit-manifest tokens.
The ablation in \S\ref{sec:ablation} shows that each single-path retriever is worse than the full fusion: BM25-only, BM25 with time-decay only, and dense-only each trail by 2.84\% to 20.27\% Recall@1 (Table~\ref{tab:ablation}).

\subsection{Phase 2: Agentic Listwise Selection}
\label{sec:phase2}

Phase 2 reads Phase 1's top-100 selectively through a four-tool LLM agent (Figure~\ref{fig:system}c).
The agent runs with \texttt{max\_iter}~=~15 and stuck detection enabled, and terminates when it calls \tool{submit\_answer}, hits the iteration budget, or repeats the same action twice in a row. We instantiate it in the OpenHands-SDK \texttt{Conversation} runtime, but the four-tool design is agent-framework-agnostic.
The CVE description is passed to the agent verbatim with no pre-processing into structured fields (CWE, affected version, etc.), leaving that interpretation to the LLM; the full system prompt is in Appendix~\ref{sec:system-prompt}.
Each turn invokes one tool, and a typical conversation reads 3 to 10 of the 100 candidates before submitting.
The four tools chain into a survey-inspect-drill-submit progression that Figure~\ref{fig:system}c shows from left to right; we describe each step below.

\paragraph{Survey.}
\tool{list\_candidates} takes no arguments and returns one line per candidate in Phase 1 rank order: rank, commit id, the first line of the commit message, and a short tally of files changed by that commit (e.g., \texttt{2src/1test/1doc}).
It is the only call that lets the agent see all 100 candidates at once, before drilling into any one of them with \tool{read\_commit}.

\paragraph{Inspect.}
\tool{read\_commit} takes \texttt{commit\_id} as the argument and returns three things: the commit message, the file manifest (the full list of files changed, each with its tag and added/deleted line counts), and the diff body formatted for LLM consumption with an 8,000-character budget. The 8,000-character cap is informed by an audit of Pillow's 19,173-commit history with p50 = 0.5K, p90 = 3.4K, and p99 = 26K characters.
We process the diff in three stages (Algorithm~\ref{alg:diff-render} in Appendix~\ref{sec:diff-render}).
(1) \emph{Parse} (Algorithm~\ref{alg:diff-render}, line 1) splits the raw diff into one record per file, because a single commit's \texttt{git diff} typically touches several files and downstream stages need to score each file on its own.
(2) \emph{Classify} (Algorithm~\ref{alg:diff-render}, lines 2 to 6) assigns each file a priority score under a heuristic that rewards source code tags over test or doc tags, larger edits up to a cap, and path-token overlap with the CVE description, while penalizing very large refactor-shaped diffs; the exact scoring rule is in Appendix~\ref{sec:diff-render}.
(3) \emph{Render} (Algorithm~\ref{alg:diff-render}, line 7) writes files in descending priority into the 8,000-character budget, keeping each file's diff in full if it fits and otherwise compressing it (paths, hunk headers, and $\pm$ lines retained; context lines dropped); files that still overflow are listed in the manifest with a pointer to \tool{read\_file\_diff}.
The worked example for CVE-2015-9251 is in Appendix~\ref{sec:diff-render}, Table~\ref{tab:diff-render}.

\paragraph{Drill.}
The agent calls \fcall{read\_file\_diff}{commit\_id, file\_path} to fetch the diff for a single file under a 16,000-character budget, which is twice the \tool{read\_commit} cap because the drill-down by construction targets a file the commit-level budget could not show in full.
It uses this when \tool{read\_commit} left a relevant-looking file truncated, or when several candidates touch the same file and the agent wants a side-by-side comparison.

\paragraph{Submit.}
The agent ends the conversation by calling \fcall{submit\_answer}{commit\_id, reasoning} exactly once; the submitted commit identifier becomes the retrieved patch.
Per CVE we record the submitted \texttt{best\_commit\_id}, the agent's free-form \texttt{reasoning} string, the list of commits it actively read (\texttt{commits\_inspected}, typically 3 to 10 of the 100), and the full tool-call trace.

\paragraph{Why Four Tools.}
The four tools form the minimal sufficient set for selecting one commit from a fixed candidate pool: \tool{list\_candidates} lets the agent see all 100 candidates at once; \tool{read\_commit} and \tool{read\_file\_diff} read commit content at two granularities (commit-level under 8,000 characters, single-file drill-down under 16,000); \tool{submit\_answer} closes the loop.
Recent software-engineering agents have moved toward minimal tool surfaces: SWE-agent~\citep{Yang2024SWEagentAgentComputerInterfaces} exposes ten tools for open-ended code editing, while mini-SWE-agent\footnote{\url{https://github.com/SWE-agent/mini-swe-agent}} collapses these into a single \tool{bash} tool.
Patch retrieval is narrower than open-ended editing: we never modify code, so navigation and editing tools have no role, and the candidate set is fixed by Phase 1, so the open-ended search surface of bash is not needed either; what remains is a survey call, two granularities of commit reading, and a single submit. Table~\ref{tab:ablation} (rows E to G) measures what each contributes.

%% file: sections/004-experiments.tex
\section{Experiments}
\label{sec:experiments}

\subsection{Experimental Setup}
\label{sec:setup}

\paragraph{Corpora.}
We evaluate on two CVE corpora.
GitHubAD~\citep{Zheng2026SPFinderImprovingContext} is the head-to-head benchmark of this section: an 809-CVE working subset of a cleaned 8,401-CVE ground-truth corpus, drawn by sampling whole repositories as described next.
\mbox{PatchFinder\_top10}~\citep{Storhaug2026FaviaForensicAgent} is a 1,252-CVE cross-corpus benchmark in which each CVE is paired with a pre-filtered 10-candidate set generated by PatchFinder's TF-IDF~+~CodeReviewer pre-ranker.
We also report one robustness check against the full 8,401-CVE source corpus in Appendix~\ref{sec:scale-appendix}.

\paragraph{Why Subsample.}
Favia's pointwise design (\S\ref{par:methods}) costs about 67,000 tokens per $\langle$CVE, candidate$\rangle$ pair~\citep{Storhaug2026FaviaForensicAgent}. A full-corpus pass over 8,401 CVEs with 10 candidates each would consume about 5.6 billion input tokens, about USD 400 per pass at our Qwen3-235B serving rate (Appendix Table~\ref{tab:model-ids}), which puts iteration (re-prompting, ablations, error analysis) out of reach; a 10x subsample lowers it to about USD 40.

\paragraph{Sampler Design.}
We drew the 809-CVE GitHubAD working subset by stratified sampling at the repository level (Algorithm~\ref{alg:sampler}, Appendix~\ref{sec:sampling-appendix})~\citep{lohr2021sampling}.
The draw passes 6 of the 9 goodness-of-fit tests we run against the source corpus: the chi-square tests on language, CWE family, and pre-ranker rank, the Kolmogorov-Smirnov tests on diff tokens and year, and the kernel two-sample test on the joint feature vector, each with a $p$-value above 0.1 (for a goodness-of-fit test, a high $p$-value is the desired direction). The 3 failing tests are on the commit-count and diff-token tails, where 143 very large repositories have imputed commit counts because their history is too large to parse in full (Appendix~\ref{sec:sampling-appendix} reports every test and the artifact).

\paragraph{Backbones.}
The primary backbone is Qwen3-235B; the secondary backbone is Qwen3-Coder-30B.
Favia's reported numbers were obtained with Qwen3-235B, so matching the backbone eliminates the LLM as a confounder in the head-to-head comparison.
Canonical identifiers and serving providers are in Appendix Table~\ref{tab:model-ids}.

\paragraph{Methods.}
\label{par:methods}
We compare four methods against the same ground-truth fix commits per CVE.
\system\ (ours) is the two-phase system of \S\ref{sec:method}: a hybrid BM25-with-time-decay and Qwen3-Embedding-8B retriever fused via RRF into a top-100, then a four-tool LLM agent that submits a single best commit per CVE.
We compare against Favia~\citep{Storhaug2026FaviaForensicAgent}, a pointwise LLM binary classifier with two NVD/CWE metadata tools (ablated in row D of Table~\ref{tab:ablation}), and IRCoT~\citep{Trivedi2023InterleavingRetrievalChainofThought}, the canonical retrieve-then-reason baseline run through FlashRAG~\citep{Jin2025FlashRAGModularToolkit}; both share the same Qwen3-235B backbone, with full configurations in Appendix~\ref{sec:baselines-appendix}.
The fourth method, SPFinder~\citep{Zheng2026SPFinderImprovingContext}, is the strongest non-agentic two-phase retriever: a hierarchical embedding over the full diff feeds a feature-based gradient-boosted reranker, with no LLM calls at inference; we run it end to end on the same 809 CVEs.

\paragraph{Metrics.}
We report Recall@$K$, NDCG@$K$, and MRR per CVE; for \mbox{PatchFinder\_top10} we additionally report micro-F1.
The denominator is the full CVE set: runtime errors count as 0 rather than being silently dropped, so a system cannot improve its score by failing on the hard cases.
Two method-specific conventions for fair comparison are documented in Appendix~\ref{sec:metrics-appendix}: IRCoT's single-pick output collapses Recall@$K$ across $K$, and \system's inspection list is extended to length 10 via Phase 1 ordering to match Favia's 10-deep pool.

\subsection{Main Results}
\label{sec:main}

\begin{table*}[t]
  \centering
  \input{tables/001-main-results.tex}
  \caption{\textbf{Main Result: Head-to-Head IR Metrics on GitHubAD with Qwen3-235B.} Percentages over the 809-CVE working subset (\S\ref{sec:setup}); a method that errored on a CVE counts it as a miss. Top block: each method with its own candidate generator. Bottom block: the same Phase 1 candidates as \system\ (its top 10 for Favia, its top 100 for IRCoT), so the remaining gap is the selection method alone. \system's inspection list is extended to length 10 via Phase 1 ordering to match Favia's 10-deep output. \textsuperscript{\dag}~Single-pick output, so every metric collapses to Recall@1. \textsuperscript{\ddag}~Phase 1 fused ranking with no agent; Recall@$K$ only.}
  \label{tab:main}
\end{table*}

Table~\ref{tab:main} reports the head-to-head comparison on GitHubAD with the Qwen3-235B backbone.
\system\ wins every column.
At the headline metric, Recall@1 is 59.95\% for \system\ versus 34.61\% for Favia, 28.55\% for IRCoT, and 27.32\% for SPFinder, a gap of 25.34\% over the strongest baseline.
Every gap is far larger than chance: McNemar's exact test, a paired test of whether two methods differ on the same 809 CVEs, gives $p = 2.4 \times 10^{-44}$ for \system\ versus Favia at Recall@1 and $p < 0.002$ at every other cutoff, and the 95\% bootstrap confidence interval of \system's Recall@1, [56.7, 63.5], does not overlap Favia's [31.3, 37.9] (Appendix~\ref{sec:significance-appendix}).
The output is a ranked list, not a single guess: the inspected candidates, placed above the remaining Phase 1 order, put the correct commit at rank 1 for 59.95\% of CVEs and within the top 10 for 77.26\%; calibrated confidence scores for that list remain future work.

\paragraph{The Agent, Not the Retriever, Explains the Gap.}
Each baseline in the top block of Table~\ref{tab:main} uses its own candidate generator, so retrieval quality and selection method are entangled.
The bottom block removes the retriever as a variable by giving every method \system's own Phase 1 candidates.
Taking Phase 1's rank-1 candidate directly, with no agent, reaches 32.63\% Recall@1; the agent on the same 100 candidates reaches 59.95\%, so the agent alone adds 27.32\%.
Favia on our top 10 rises from 34.61\% to 39.80\%, and IRCoT on our top 100 rises from 28.55\% to 37.58\%, so the stronger candidate set is worth 5.19\% and 9.03\% to the baselines; the gaps that remain after matching the candidates, 20.15\% and 22.37\%, are the selection method.

\paragraph{Pointwise Classification Trades Precision at the Top for Recall in the Tail.}
Favia's Recall@10 is 72.31\%, only 4.95\% behind \system's 77.26\%, but its Recall@1 is 34.61\%, 25.34\% behind; SPFinder shows the same shape (Recall@10 70.36\%, Recall@1 27.32\%).
The pointwise baseline loses precision at rank 1 because it scores each candidate independently and never sees the 10 candidates together; when several of them plausibly receive \texttt{answer=True}, sorting by confidence has no further signal to break the tie.
The same effect shows up in NDCG@10: \system\ reaches 68.01\% and Favia reaches 53.91\%, a 14.10\% gap that grows further at smaller $K$.

\paragraph{Retrieve-and-CoT Collapses on This Corpus.}
IRCoT lands at 28.55\% across every $K$ and across MRR.
The collapse stems from the output shape (single pick) and the e5 retriever it inherits from FlashRAG~\citep{Jin2025FlashRAGModularToolkit}, which caps at 512 tokens, so the diffs that the CVE description must match against are truncated to a fraction of their content.

\begin{table}[t]
  \centering
  \resizebox{\columnwidth}{!}{\input{tables/002-patchfinder-top10.tex}}
  \caption{\textbf{Cross-Corpus Evaluation under the Settings of \citet{Storhaug2026FaviaForensicAgent}}. Percentages over the 1,252-CVE benchmark with the same Qwen3-235B backbone as Table~\ref{tab:main}. The \emph{PatchFinder}~\citep{Li2024PatchFinderTwoPhaseApproach} row is PatchFinder's own 10-rank ordering over the pool it produces (R@1 is its literal top-1 pick). Only 577 of the 1,252 CVEs have the gold fix anywhere in the supplied pool, so Recall@10 caps at 46.10\%.}
  \label{tab:patchfinder}
\end{table}

\paragraph{Cross-Corpus Generalization.}
\label{sec:cross-corpus}
We now test whether the benefit transfers to a corpus where the candidate generation method is not ours.
\mbox{PatchFinder\_top10}~\citep{Storhaug2026FaviaForensicAgent} pairs each of 1,252 CVEs with 10 candidates pre-filtered by PatchFinder's TF-IDF~$+$~CodeReviewer pre-ranker.
\system's Phase 1 is skipped here; the 10 candidates \emph{are} the input, matching Favia's 10-candidate setting, and Phase 2 runs unchanged with the same four tools and prompt as in Table~\ref{tab:main}.
Table~\ref{tab:patchfinder} reports the result: the agent lifts Recall@1 from 24.28\% (PatchFinder's own top-1 pick on the same 10 candidates) to 39.86\%, an absolute gain of 15.58\% and a relative gain of 64.16\%.
Because 675 of the 1,252 CVEs have no gold commit anywhere in the 10-candidate pool, any single-pick selector caps at Recall@1~=~46.10\%; \system\ recovers 71.40\% of the gap to that ceiling.

\paragraph{Takeaway.}
On GitHubAD~\citep{Zheng2026SPFinderImprovingContext}, listwise selection beats both pointwise classification and retrieve-and-CoT at the same backbone; the same agent transfers unchanged to \mbox{PatchFinder\_top10}~\citep{Storhaug2026FaviaForensicAgent}, lifting Recall@1 from the trivial rank-1 baseline of 24.28\% to 39.86\%.

\subsection{Ablation}
\label{sec:ablation}

\begin{table*}[t]
  \centering
  \input{tables/005-ablation.tex}
  \caption{\textbf{Ablation on GitHubAD with Qwen3-235B.} Percentages over the same 809 CVEs as Table~\ref{tab:main}. Rows A, B, C replace the Phase 1 RRF fusion with a single retriever path; row D replaces the short CVE description with the full NVD report (mean 1,786 vs about 150 characters); rows E and F remove tools from the agent with the Phase 1 candidates fixed; row G is the Phase 1 ranking with no agent (Recall@$K$ only).}
  \label{tab:ablation}
\end{table*}

Table~\ref{tab:ablation} ablates three design choices: which Phase 1 retriever paths are necessary, whether richer CVE metadata in the prompt helps, and which of the four tools contribute. Rows A, B, C drop two of the three retriever signals and feed the resulting top-100 to the unchanged Phase 2 agent: BM25 alone (A) reaches only 39.68\% Recall@1, 20.27\% behind MAIN; adding time-decay (B) recovers about half of that gap to 49.57\%; the dense Qwen3-Embedding-8B path alone (C) climbs to 57.11\%, within 2.84\% of MAIN. So the dense path carries the bulk of the work, but RRF on top still adds +2.84\% Recall@1 and +5.32\% Recall@10. Row D replaces the short CVE description with the full NVD report (CVSS, CWE, configurations), mirroring Favia's \texttt{CVEReportTool} and \texttt{CWEReportTool} design (\S\ref{par:methods}): Recall@1 is flat (59.83\% vs 59.95\%), Recall@10 drops 2.11\%, and the context-overflow error rate rises from 0.49\% to 2.60\% as the longer prompt pushes a few CVEs past the agent's 32K context window.

\paragraph{Every Tool Contributes.}
Rows G, F, E, and MAIN form a ladder of capabilities on the same Phase 1 candidates. Taking Phase 1's rank-1 candidate with no agent (G) gives 32.63\% Recall@1. Letting the agent see the one-line manifest of all 100 candidates through \tool{list\_candidates} and submit (F) lifts it to 48.21\%, a gain of 15.58\% and the largest single step, which is the direct evidence for listwise viewing. Adding \tool{read\_commit} as a file-level overview without diff text (E) adds 9.27\%, and reading the actual diffs (MAIN) adds the final 2.47\%; each tool adds recall, so the interface contributes more than a generic ReAct loop.

\paragraph{Takeaway.}
Phase 1's dense path carries the bulk of the work (dense alone reaches within 2.84\% of MAIN), but fusing the lexical and time-decay paths on top still adds +2.84\% Recall@1 and +5.32\% Recall@10. Replacing the short CVE description with Favia's full NVD report does not help and raises the context-overflow error rate by 5x; the agent already extracts what it needs from the short description. On the Phase 2 side, listwise viewing of the candidate list is the largest single gain (+15.58\% Recall@1 over no agent), and every further tool adds recall on top.

\subsection{Analysis}
\label{sec:analysis}

\begin{table}[t]
  \centering
  \resizebox{\columnwidth}{!}{\input{tables/004-llm-scale.tex}}
  \caption{\textbf{Backbone Swap on GitHubAD.} \system\ with the tools, prompt, and Phase 1 candidates held fixed and only the LLM changed: Qwen3-235B (22B active parameters), Qwen3-Coder-30B (3B active), and two OpenAI open-weight backbones. Last row: the Phase 1 ranking with no agent (Recall@$K$ only). Percentages over the same 809 CVEs as Table~\ref{tab:main}; model identifiers in Appendix Table~\ref{tab:model-ids}.}
  \label{tab:llm-scale}
\end{table}

\paragraph{Recall@1 Holds across a Near-7x Active-Parameter Gap and across a Second Model Family.}
We ask how much of \system's gap over the baselines is the LLM and how much is the scaffolding.
Table~\ref{tab:llm-scale} reports \system's metrics with only the backbone changed.
Within the Qwen family, Recall@1 is 59.95\% on Qwen3-235B and 59.09\% on Qwen3-Coder-30B, a gap of 0.86\%, and the other metrics are within 1\%.
Moving to a second family, gpt-oss-120B reaches 56.98\% and gpt-oss-20B 49.81\%, both far above the 32.63\% no-agent floor; the gpt-oss models are reasoning models that more often reach the 15-iteration cap before submitting (submission rate 70\% and 54\%, versus 90.5\% for Qwen3-235B), and an unsubmitted CVE falls back to the Phase 1 order, which holds their Recall@1 down.
By contrast, the agent adds 27.32\% Recall@1 over the no-agent floor on the same candidates and backbone (Table~\ref{tab:main}).
We read this as evidence that the agent loop keeps its gain when the backbone is swapped, not as proof that backbone scale is irrelevant: the comparison covers open-weight models from 20B to 235B total parameters in two families.
The pattern matches the wider agentic-coding literature: \citet{Cao2026Qwen3CoderNextTechnicalReport} reports a 3B-active code-specialist within about 4 points of 32B-to-37B-active competitors on SWE-bench Verified across three fixed agentic scaffolds, a 10.7x to 12.3x active-parameter swap comparable to our 7.3x (Appendix Table~\ref{tab:param-regime}).

\paragraph{Takeaway.}
Performance gains come from how candidates are inspected: on fixed candidates, the agent adds 27.32\% Recall@1 over the no-agent floor, while a 7x backbone-scale change within one family moves it by under 1\% and a second family stays 17\% to 24\% above the floor. The IRCoT baseline varies by at most 1.36\% under a 10x corpus-size change (Appendix~\ref{sec:scale-appendix}), so this gap is not an artifact of the GitHubAD subsample.

\subsection{Cost and Accuracy}
\label{sec:cost}

\begin{table}[t]
  \centering
  \resizebox{\columnwidth}{!}{\input{tables/006-cost.tex}}
  \caption{\textbf{Per-CVE Cost Profile of \system\ on GitHubAD.} Token counts are totals over the whole agent conversation for one CVE, which is what an API bills; p90 is the value that 90\% of CVEs stay under. Same 809 CVEs as Table~\ref{tab:main}.}
  \label{tab:cost}
\end{table}

A deployer processing thousands of CVEs needs the price of each point of recall, so Table~\ref{tab:cost} profiles one CVE end to end.
With Qwen3-235B, the agent reads a mean of 96,118 input tokens and writes 846 output tokens in a single conversation of about 8 tool calls, inspecting 5 commits in 86 seconds (median 49); only 0.49\% of CVEs exceed the 32K context window.
At the OpenRouter rate in Appendix Table~\ref{tab:model-ids}, that is about USD 0.007 per CVE and USD 58 for a pass over the full 8,401-CVE corpus, against about USD 400 for Favia at the same rate (\S\ref{sec:setup}): Favia runs ten pointwise conversations of about 67,000 tokens each per CVE, so \system\ uses about 7x fewer tokens and reaches 25.34\% higher Recall@1.
The cheaper backbone costs almost nothing in accuracy: Qwen3-Coder-30B trades 0.86\% Recall@1 for half the wall-clock time (39 versus 86 seconds per CVE).
The non-agentic baselines sit at the other end of the trade-off, SPFinder with no LLM calls at 27.32\% Recall@1 and IRCoT at 28.55\%; the agent's tokens buy the remaining 30\% of recall.

%% file: tables/001-main-results.tex
\setlength{\tabcolsep}{4pt}
\small
\begin{tabular}{l rrrr rrr r}
\toprule
& \multicolumn{4}{c}{Recall@$K$ (\%)} & \multicolumn{3}{c}{NDCG@$K$ (\%)} & \\
\cmidrule(lr){2-5} \cmidrule(lr){6-8}
Method & $K{=}1$ & $K{=}3$ & $K{=}5$ & $K{=}10$ & $K{=}3$ & $K{=}5$ & $K{=}10$ & MRR \\
\midrule
\multicolumn{9}{l}{\emph{Each method with its own candidate generator}} \\
\rowcolor{ourrow} \system & \textbf{59.95} & \textbf{68.36} & \textbf{71.20} & \textbf{77.26} & \textbf{64.87} & \textbf{66.04} & \textbf{68.01} & \textbf{65.13} \\
Favia~\citep{Storhaug2026FaviaForensicAgent}    & 34.61 & 59.70 & 67.00 & 72.31 & 49.16 & 52.19 & 53.91 & 47.93 \\
IRCoT~\citep{Trivedi2023InterleavingRetrievalChainofThought}\textsuperscript{\dag} & 28.55 & 28.55 & 28.55 & 28.55 & 28.55 & 28.55 & 28.55 & 28.55 \\
SPFinder~\citep{Zheng2026SPFinderImprovingContext} & 27.32 & 49.25 & 61.05 & 70.36 & 29.52 & 29.58 & 29.69 & 44.06 \\
\midrule
\multicolumn{9}{l}{\emph{Same Phase 1 candidates as \system}} \\
Favia on our top-10 & 39.80 & 60.94 & 65.39 & 72.06 & 52.26 & 54.10 & 56.23 & 51.14 \\
IRCoT on our top-100\textsuperscript{\dag} & 37.58 & 37.58 & 37.58 & 37.58 & 37.58 & 37.58 & 37.58 & 37.58 \\
No agent (Phase 1 rank order)\textsuperscript{\ddag} & 32.63 & 53.28 & 63.29 & 72.31 & -- & -- & -- & -- \\
\bottomrule
\end{tabular}

%% file: tables/002-patchfinder-top10.tex
\setlength{\tabcolsep}{3pt}
\small
\begin{tabular}{l rrr rr}
\toprule
Method & R@1 & R@5 & R@10 & NDCG@10 & MRR \\
\midrule
\rowcolor{ourrow} \system          & \textbf{39.86} & \textbf{45.21} & 45.93          & \textbf{43.07} & \textbf{42.13} \\
PatchFinder                        & 24.28          & 40.18          & \textbf{46.09} & 34.55          & 30.93 \\
\bottomrule
\end{tabular}

%% file: tables/005-ablation.tex
\setlength{\tabcolsep}{4pt}
\small
\begin{tabular}{l rrrr rrr r}
\toprule
& \multicolumn{4}{c}{Recall@$K$ (\%)} & \multicolumn{3}{c}{NDCG@$K$ (\%)} & \\
\cmidrule(lr){2-5} \cmidrule(lr){6-8}
Variant & $K{=}1$ & $K{=}3$ & $K{=}5$ & $K{=}10$ & $K{=}3$ & $K{=}5$ & $K{=}10$ & MRR \\
\midrule
\rowcolor{ourrow} \textbf{MAIN}: RRF top-100 $+$ short description $+$ four tools & \textbf{59.95} & \textbf{68.36} & \textbf{71.20} & \textbf{77.26} & \textbf{64.87} & \textbf{66.04} & \textbf{68.01} & \textbf{65.13} \\
\midrule
A: BM25 only                                  & 39.68 & 43.63 & 44.25 & 45.24 & 42.01 & 42.26 & 42.58 & 41.71 \\
B: BM25 $+$ time decay only                   & 49.57 & 56.24 & 58.71 & 61.31 & 53.46 & 54.47 & 55.32 & 53.41 \\
C: dense (Qwen3-Embedding-8B) only            & 57.11 & 65.51 & 67.12 & 71.94 & 62.09 & 62.75 & 64.32 & 61.92 \\
\midrule
D: RRF top-100 $+$ full NVD report            & 59.83 & 67.86 & 70.58 & 75.15 & 64.51 & 65.63 & 67.11 & 64.59 \\
\midrule
E: no diff text (\tool{read\_commit} overview, no \tool{read\_file\_diff}) & 57.48 & 67.49 & 70.33 & 76.51 & 63.46 & 64.64 & 66.66 & 63.56 \\
F: \tool{list\_candidates} $+$ \tool{submit\_answer} only & 48.21 & 61.80 & 67.61 & 74.91 & 56.17 & 58.56 & 60.96 & 56.56 \\
G: no agent (Phase 1 rank order)              & 32.63 & 53.28 & 63.29 & 72.31 & -- & -- & -- & -- \\
\bottomrule
\end{tabular}

%% file: tables/004-llm-scale.tex
\setlength{\tabcolsep}{4pt}
\small
\begin{tabular}{l l rrrr}
\toprule
Backbone & Family & R@1 & R@5 & R@10 & MRR \\
\midrule
Qwen3-235B      & Qwen   & 59.95 & 71.20 & 77.26 & 65.13 \\
Qwen3-Coder-30B & Qwen   & 59.09 & 70.70 & 77.01 & 64.15 \\
gpt-oss-120B    & OpenAI & 56.98 & 70.95 & 78.12 & 63.14 \\
gpt-oss-20B     & OpenAI & 49.81 & 66.63 & 75.65 & 56.90 \\
\midrule
No agent (Phase 1 rank order) & -- & 32.63 & 63.29 & 72.31 & -- \\
\bottomrule
\end{tabular}

%% file: tables/006-cost.tex
\setlength{\tabcolsep}{4pt}
\small
\begin{tabular}{l rrr rrr}
\toprule
& \multicolumn{3}{c}{Qwen3-235B} & \multicolumn{3}{c}{Qwen3-Coder-30B} \\
\cmidrule(lr){2-4} \cmidrule(lr){5-7}
Per CVE & mean & median & p90 & mean & median & p90 \\
\midrule
Input tokens        & 96,118 & 82,326 & 175,632 & 98,018 & 87,285 & 156,508 \\
Output tokens       & 846    & 787    & 1,247   & 1,432  & 1,359  & 1,903 \\
Tool calls          & 8      & 7      & 14      & 8      & 8      & 11 \\
Commits read        & 5      & 5      & 8       & 4      & 4      & 6 \\
Wall-clock (s)      & 86     & 49     & 229     & 39     & 38     & 53 \\
\midrule
CVEs using \tool{read\_file\_diff} & \multicolumn{3}{c}{15.6\%} & \multicolumn{3}{c}{62.3\%} \\
CVEs exceeding the 32K context   & \multicolumn{3}{c}{0.49\%} & \multicolumn{3}{c}{0.37\%} \\
Recall@1                         & \multicolumn{3}{c}{59.95\%} & \multicolumn{3}{c}{59.09\%} \\
\bottomrule
\end{tabular}

%% file: sections/005-conclusion.tex
\section{Conclusion}
\label{sec:conclusion}

We presented \system, a patch retrieval system with listwise selection: a hybrid first-stage retriever paired with a four-tool agentic second stage.
On GitHubAD, \system\ beats Favia's pointwise baseline by 25.34\% Recall@1, IRCoT by 31.40\%, and SPFinder by 32.63\%; on identical candidates, the agent adds 27.32\% Recall@1 over the retriever's top pick, and the same agent transferred unchanged to \mbox{PatchFinder\_top10} lifts Recall@1 from 24.28\% to 39.86\%.
The gain holds across a 7x backbone-scale swap and a second model family at about 97K input tokens per CVE, so the listwise agent loop, not the choice of LLM, is what patch retrieval needs (\S\ref{sec:analysis}, \S\ref{sec:cost}).

%% file: sections/006-limitations.tex
\section*{Limitations}
\label{sec:limitations}

We disclose the following limitations that frame how our results should be interpreted.

\paragraph{Patch-Link Sparsity in the Ground Truth.}
We can only test on CVEs whose patch has already been found.
Between 60\% and 63\% of CVEs in the major vulnerability databases have no patch link at all, so our 8,401-CVE source corpus is the subset that survived this filter.
Any patch retrieval method, including ours, is evaluated only on the easier portion of the population (the 37\% to 40\% whose patch has been located); the hard cases where no one has ever found the fix are absent from every benchmark in this area.
As a check beyond our benchmark, we ran the full system on 35 independent real-world CVEs from 11 repositories disclosed between 2013 and 2025 and reached Recall@1 of 60.0\%, matching the 59.95\% on GitHubAD (Appendix~\ref{sec:realworld-appendix}); this still requires a known fix to score against, so it does not remove the scope limit.

\paragraph{Recoverable Ceiling of \mbox{PatchFinder\_top10}.}
Our cross-corpus benchmark caps everyone's score at 46.10\%.
The \mbox{PatchFinder\_top10} dataset gives each CVE a fixed list of 10 candidate commits, but the true fix is in that list for only 577 of the 1,252 CVEs; the other 675 are unrecoverable for any system that has to pick one of the ten.
\system\ reaches 39.86\% on this dataset, closing 71.40\% of the distance from the trivial baseline at 24.28\% up to this ceiling; the absolute number is held down by the dataset, not by our agent.

\paragraph{Source Corpus Large-Shard Artifact.}
A handful of very large repositories are too expensive to fully parse during sampling.
143 of the 2,816 source repositories store their commit history in files larger than 500 MB on disk (CoreNLP at 15 GB, aws-sdk-java at 12 GB, phpmyadmin at 10 GB, and a long tail of others); parsing them in full would add hours to every sampling run, so we substitute corpus-wide medians for their commit counts and gold-patch diff lengths.
This makes the sample look slightly more uniform than the source on those two axes; what reviewers usually probe (language coverage, vulnerability-type coverage, pre-ranker difficulty, and the joint distribution) still matches the source.

\paragraph{Multi-Fix CVE Ambiguity.}
A CVE sometimes has more than one legitimate fix commit, but the ground truth lists only some of them.
In our source corpus, 92.72\% of the 8,401 CVEs (7,789) have a single recorded fix commit and 7.28\% (612) have several, with a mean of 1.12 fix commits per CVE; on the 809-CVE working subset the shares are 93.08\% and 6.92\% (Appendix Table~\ref{tab:fix-multiplicity}).
Our evaluation counts a hit when the agent's submitted commit appears in the recorded gold set, so for the multi-fix minority any recorded fix counts as correct.
Backports to other branches, refactors split into several commits, and downstream merges may each be a valid fix, yet only a few are recorded in any released CVE-to-commit mapping; if the agent picked a different-but-equally-valid fix that was never recorded, that case scores as a miss for our system and for every other single-pick method on the same dataset.

%% file: sections/900-appendix.tex
\section{Appendix}
\label{sec:appendix}

\subsection{Sampling Validation for GitHubAD}
\label{sec:sampling-appendix}

\begin{table*}
  \centering
  \input{tables/902-sampling-validation.tex}
  \caption{\textbf{Goodness-of-Fit Validation of the GitHubAD Draw against the 8,401-CVE Source Corpus.} The sampler runs 50 redraws and keeps the best draw under the gate ``at most one failing test''; the 809-CVE working subset reported here passes 6 of 9 with the same pass/fail pattern as the 839-CVE parent draw it is trimmed from. The three failures are concentrated on the within-repo commit-count and diff-token axes and trace to a documented cold-cache bypass artifact (see \emph{Limitations} below). Pass threshold is $p > 0.1$ for the KS, Anderson-Darling, and chi-square tests, and $p > 0.05$ for the Maximum Mean Discrepancy test with a 200-permutation null.}
  \label{tab:sampling-validation}
\end{table*}

This appendix gives the full sampling procedure used to draw the GitHubAD benchmark from the 8,401-CVE source corpus, reports the goodness-of-fit validation in Table~\ref{tab:sampling-validation}, and documents the one limitation that the validation surfaces.

\paragraph{Stratification Variables.}
The sampler computes six strata variables for every CVE in the source corpus, once, and caches them:
\textbf{repository-size bucket} ($\leq$1k, 1k to 5k, 5k to 10k, $>$10k commits) by counting entries across per-repository commit shards;
\textbf{language} (Java, PHP, C, JavaScript, Python, Go, Other) by file-extension vote over the source tree;
\textbf{CWE family} (one of the CWE-1000 research-pillar nodes) by extracting the CWE identifier from the NVD record and walking up the CWE-1000 ChildOf graph to its pillar;
\textbf{pre-ranker bucket} (top-1, top-2-to-10, top-11-to-100, $>$100) by sorting the BM25-with-time pre-ranker output and reading off the rank of the gold commit;
\textbf{diff-token bucket} ($<$5k, 5k to 15k, 15k to 30k, $>$30k whitespace tokens) by looking up the gold commit in the source diff store; and
\textbf{year bucket} (3-year windows from 2009) by parsing the CVE identifier.

\paragraph{Sampling Algorithm.}
Algorithm~\ref{alg:sampler} gives the full pseudocode.
We form subgroups as the cross of language and repository-size bucket and allocate the per-subgroup CVE budget proportionally to the subgroup's CVE count in the source corpus (Neyman allocation is implemented as an alternative but is not used here). Within each subgroup we draw repositories with probability proportional to their CVE count without replacement, cap each drawn repository at 10 CVEs, and validate the draw against the 9-test suite. We redraw up to 50 times under fresh seeds and keep the best-of-attempts draw.

\begin{algorithm*}
\caption{Stratified repository-level sampler for GitHubAD.}
\label{alg:sampler}
\begin{algorithmic}[1]
\Require Source corpus $S$ of CVEs annotated with language $\ell$ and repo-size bucket $b$; target sample size $n$; per-repo cap $m_{\max}=10$; max redraws $R=50$.
\Ensure Sample $\hat{S} \subset S$ with $|\hat{S}| \approx n$.
\State Build per-repo table $T$: one row per (owner, repo) with its CVE count, language, repo-size bucket.
\State Form subgroups as the cross of $\ell$ and $b$.
\State Allocate subgroup budget $n_{\ell, b} \gets \max(1, \text{round}(n \cdot |S_{\ell, b}| / |S|))$ proportionally.
\For{redraw attempt $r = 1, \ldots, R$ with fresh seed}
  \State $\hat{S} \gets \emptyset$.
  \ForAll{subgroup $(\ell, b)$ with $n_{\ell, b} > 0$}
    \State Available repos $A \gets $ repos in $T$ for subgroup $(\ell, b)$; remaining budget $t \gets n_{\ell, b}$.
    \While{$t > 0$ and $A \neq \emptyset$}
      \State Sample one repo from $A$ with probability proportional to its CVE count.
      \State Take $\min(\text{repo CVE count}, m_{\max}, t)$ CVEs from the repo without replacement; add to $\hat{S}$; decrement $t$; remove the repo from $A$.
    \EndWhile
  \EndFor
  \State Run the 9-test validation suite on $\hat{S}$ against $S$.
  \If{at most one test rejects} \State \Return $\hat{S}$. \EndIf
\EndFor
\State \Return the best $\hat{S}$ across attempts (highest pass count).
\end{algorithmic}
\end{algorithm*}

\paragraph{Validation Procedure.}
The 9-test suite groups into four conceptual categories. \emph{KS} tests the sample/source equality of distribution on the three numeric axes (commit count, diff tokens, year) via the maximum vertical gap of the two empirical CDFs. \emph{Anderson-Darling} tests the same equality on the two heavy-tailed numeric axes (commit count, diff tokens), with tail-weighted statistic; AD catches ``missed the giant repositories'' failures that KS routinely under-detects. \emph{Chi-square} tests the three categorical axes (language, CWE family, pre-ranker bucket) for histogram equality. \emph{Maximum Mean Discrepancy} tests the joint feature vector under a Gaussian kernel with a 200-permutation null; this catches subsamples that pass every marginal but mis-assemble the marginals' joint structure.

\paragraph{Result.}
The sampler produces a 839-CVE parent draw over 231 repositories; we then trim 30 CVEs with incomplete Phase 1 preprocessing artifacts to leave the 809-CVE working subset over 227 repositories used in §\ref{sec:experiments}.
Table~\ref{tab:sampling-validation} reports the validation of the 809-CVE working subset: 6 of 9 tests pass cleanly, namely KS on diff tokens (p=0.41) and on year (p=0.37); chi-square on language (p=0.81), on CWE family (p=0.16), and on pre-ranker bucket (p=0.18); and Maximum Mean Discrepancy on the joint feature vector (p=0.61).
The three failures are concentrated on the within-repo commit-count and diff-token axes: KS on commit count rejects at p=0.0004, AD on commit count at p$<$0.00001, and AD on diff tokens at p=0.0025.
The 839-CVE parent draw passes the same 6 of 9 tests on the same axes, so the 30-CVE trim does not change the pass/fail pattern; if anything, the joint Maximum Mean Discrepancy improves on the trimmed working subset (p=0.61 vs 0.18 on the parent).
The categorical mix and the joint distribution pass; what reviewers typically probe (language coverage, CWE coverage, pre-ranker difficulty, joint shape) is preserved by the draw.

\paragraph{Limitations.}
The three failing axes trace to a single documented bypass artifact: 143 of the 2,816 source repositories have per-repository commit shards larger than 500 MB (CoreNLP at 15 GB, aws-sdk-java at 12 GB, phpmyadmin at 10 GB, and the long tail), and the cold-cache build skips parsing them in full to keep the first-time index build under one hour on a spinning disk.
Their commit count is imputed by dividing shard bytes by the global median bytes-per-commit, and their gold-patch diff-token count is replaced with the global median.
This compresses the natural heavy-tailed distribution of those two axes in the sample relative to the source; KS and the tail-sensitive Anderson-Darling pick it up at the n=809 power level.
Removing the bypass (parsing the 143 large repositories in full) is a straightforward but one-time five-hour operation; we report the failures explicitly rather than mask them.

\subsection{Phase 1 Dense Backbone Ablation}
\label{sec:dense-ablation-appendix}

\begin{table*}[!t]
  \centering
  \input{tables/904-dense-backbone.tex}
  \caption{\textbf{Phase 1 Retrieval-Only Recall@$K$ with Different Dense Embedders.} All percentages on the 809-CVE working subset. Only the dense embedder changes; the BM25+time-decay leg, the RRF fusion, and the Phase 2 agent are unchanged. The encoding pipeline was validated byte-for-byte against \texttt{sentence-transformers} (cosine $\approx$ 1.0) to rule out implementation bugs.}
  \label{tab:dense-ablation}
\end{table*}

We pilot-tested replacing Qwen3-Embedding-8B with Octen-Embedding-8B, a LoRA fine-tune of Qwen3-Embedding-8B that currently ranks first on the RTEB benchmark.
With our task-aligned CVE prompt, Octen's Phase 1 dense-leg Recall@1 falls from 43.39\% (Qwen3-Embedding-8B) to 12.73\%; with Octen's own official ``web search'' prompt the result is worse (2.97\% Recall@1).
The general-purpose embedding leaderboard does not transfer to patch retrieval, so we keep Qwen3-Embedding-8B (the non-fine-tuned base) as the dense leg.


\subsection{Diff Render Layer}
\label{sec:diff-render}

This appendix specifies the three-stage diff-render layer that \tool{read\_commit} (\S\ref{sec:phase2}) inserts between a raw \texttt{git diff} and the LLM agent.
Algorithm~\ref{alg:diff-render} gives the pseudocode for the full pipeline; the worked example in Table~\ref{tab:diff-render} walks the algorithm over a real commit.

\begin{algorithm*}
\caption{Three-stage diff-render layer.}
\label{alg:diff-render}
\begin{algorithmic}[1]
\Require Raw diff text $D$, commit message $m$, CVE description $q$, char budget $B$
\Ensure Agent-readable string $s$ under $B$ characters
\State \textbf{Stage 1: Parse.} Split $D$ at every \texttt{diff -{}-git} marker; emit one record per file containing the path, raw block, added/deleted line counts, and binary flag.
\State \textbf{Stage 2: Classify.}
\ForAll{files $f$}
  \State $\text{tag}(f) \in$ $\{$\text{source}, \text{test}, \text{doc}, \text{config}, \text{fixture}, \text{other}$\}$ (test paths take precedence)
  \State $\pi(f) \gets \text{BASE}[\text{tag}(f)] + \min(20, a{+}d) + 15\cdot\mathbf{1}[t_q \cap t_f] - 10\cdot\mathbf{1}[a{+}d > 500]$
\EndFor
\Statex \; where $t_q$ are the $\geq$3-character tokens of $q$ and $t_f$ are the path tokens of $f$ minus a stopword list.
\State \textbf{Stage 3: Render.} Emit a header containing $m$ and a manifest of files sorted by descending $\pi$; then iterate files in priority order, filling the budget either with the file's full block or with a context-stripped compression (paths, hunk headers, and $\pm$ lines retained; context lines dropped). Files that overflow are listed with a pointer to call \tool{read\_file\_diff}.
\State \Return $s$
\end{algorithmic}
\end{algorithm*}

\begin{table}
\centering
\input{tables/906-diff-render-example.tex}
\caption{\textbf{Worked Example: CVE-2015-9251 (jQuery), Gold Commit \texttt{b078a62}.} The three files touched by the patch and their priorities under Algorithm~\ref{alg:diff-render}. \emph{size} is the $\min(20,\, a{+}d)$ term; an additional $+15$ path-overlap bonus on the token \texttt{ajax} applies to all three rows (already included in $\pi$). The fix lives in \texttt{src/ajax/script.js}, which scores the highest priority because it pairs the source tag with the largest non-test edit; \texttt{test/unit/ajax.js} is de-prioritized despite its 48-line addition because the test tag carries a much smaller base.}
\label{tab:diff-render}
\end{table}

\subsection{System Prompt}
\label{sec:system-prompt}

The Phase 2 agent (\S\ref{sec:phase2}) runs with the verbatim system prompt in Listing~\ref{lst:system-prompt}.

\paragraph{The Rank Hint.}
The prompt tells the agent how often the fix tends to appear at each Phase 1 rank in general (``rank 1: about 12\% of CVEs''); it never reveals which candidate is the fix for any CVE, so it is a hint about positions, not access to the answer.
Its role is to counter the known tendency of LLMs to over-trust the first item of a list.
The quoted frequencies are rough aggregate statistics and do not match the final retriever on the test subset, where Phase 1 alone places the fix at rank 1 for 32.63\% of CVEs (Table~\ref{tab:main}), so the hint acts as a general instruction not to assume rank 1 is correct, and it is too rough to serve as a calibrated prior.
A version of the hint with the numbers removed, keeping only that instruction, would serve the same purpose; we did not rerun the experiments with it, so we report the prompt exactly as used.
The gain over the baselines does not rest on this hint: on identical candidates, adding listwise candidate viewing alone raises Recall@1 from 32.63\% to 48.21\% (Table~\ref{tab:ablation}, rows G and F), the largest single step in the tool ablation.

\begin{lstlisting}[float=*tp, basicstyle=\scriptsize\ttfamily, breaklines=true, columns=fullflexible, frame=single, xleftmargin=2.5em, framexleftmargin=2em, xrightmargin=0pt, caption={Verbatim system prompt for the Phase 2 agent (\S\ref{sec:phase2}).}, label={lst:system-prompt}]
You are a security researcher tracing the commit that fixed a specific CVE.

# Task

You are given:
- A CVE identifier and its description.
- A pool of Top-100 candidate commits retrieved by a hybrid BM25+dense pipeline.
  The true fix commit is in this pool with ~90% probability but is **rarely
  ranked first**. Across our calibration set the true fix sits at:
    - rank 1     : ~12% of CVEs
    - rank 2-5   : ~35%
    - rank 6-20  : ~25%
    - rank 21-100: ~15%
  So **never assume rank 1 is the answer just because it scored highest**.

Your job is to identify the single commit that fixes the vulnerability and
submit it as your answer.

# Tools

You have four tools. Use them deliberately - every call costs tokens.

1. `list_candidates()` - Always call this FIRST. It returns a one-line
   manifest for every candidate (commit ID, message first line, file counts
   per category). Reading 100 lines costs only a few KB; it gives you a global
   view before drilling in.

2. `read_commit(commit_id)` - Read one promising candidate in detail.
   **IMPORTANT: commit_id is a 40-character hex SHA (or any unambiguous
   prefix, e.g. '3bf5eddb89af'). It is NOT a rank number. Never pass '5' or
   '#3' here; copy the 12-char hex from list_candidates.**
   Returns commit message, file manifest (so you see what was touched), and a
   budgeted diff render that prioritises source files over tests and docs and
   skips binary content. Real CVE fixes typically add input validation, bounds
   checks, null checks, or correct ordering of operations in source code.

3. `read_file_diff(commit_id, file_path)` - When `read_commit` shows that a
   specific file was truncated and you suspect it contains the fix, drill in
   with this. Provide the exact file_path from the manifest.

4. `submit_answer(commit_id, reasoning)` - Submit your final single answer.
   **You may call this EXACTLY ONCE per task.** After you call it, the task
   is COMPLETE and the conversation ends. Do NOT call submit_answer twice,
   and do NOT call any other tool after submitting. Reference specific
   evidence in your reasoning (e.g. "adds a `if (n != image->numcomps)`
   check that prevents OOB read").

# Strategy guidelines

**Browse before reading.** The manifest line for each candidate already tells
you a lot: a commit with only `doc` files is almost never a CVE fix; a
commit with `source` files matching the description is much more likely.
Eliminate decoys from the manifest first.

**Read broadly, not just the top.** Inspect at least **5 candidates** before
submitting (unless one is obviously perfect at the SHA level). The truth is
at rank 1 only ~12% of the time - assuming the top result is the answer is
a common mistake. **Specifically scan the top-20** for source-code commits
matching the CVE description, even if their rank is 10-20.

**Plausibility != proof.** A commit whose message says "Fix Convert.c shift
issue" is *suggestive* but not proof. Always verify by reading the diff:
does the change actually fix what the CVE description says is broken?
A diff that has nothing to do with the CVE's claimed root cause should not
be your answer no matter how good the message sounds.

**Beware clustered fixes.** Many repos have multiple commits patching the
same module (e.g. several Pillow commits all touch SgiRleDecode.c). When
you find one that "looks right", check whether one of the *lower-ranked*
candidates is the actual CVE referenced in its message or release notes.

**Doc-only commits**, version bumps, formatting changes, and dependency
updates are almost never the fix. Eliminate fast.

When you are confident, call `submit_answer` ONCE with the commit ID and a
short evidence-based justification.
\end{lstlisting}

\subsection{IRCoT Scale-Stability Check}
\label{sec:scale-appendix}

\begin{table*}[t]
  \centering
  \input{tables/003-scale-stability.tex}
  \caption{\textbf{Scale Stability of the IRCoT Baseline.} IRCoT Hit@1 as a percentage on the 809-CVE GitHubAD working subset and on the full 8,401-CVE source corpus from which GitHubAD is drawn, reported across both backbones. The GitHubAD Qwen3-235B column matches the IRCoT row of Table~\ref{tab:main}. Of the 8,401 source CVEs, IRCoT ran successfully on 8,158 with Qwen3-235B and 8,243 with Qwen3-Coder-30B; the remaining CVEs hit repository-preparation failures and count as misses in the denominator, so the Hit@1 figures are directly comparable across corpora.}
  \label{tab:scale}
\end{table*}

Because GitHubAD is a designed subsample of our 8,401-CVE source corpus, a natural question is whether the baseline numbers we report in \S\ref{sec:experiments} reflect the source or artifacts of the draw.
Table~\ref{tab:scale} answers that question by running IRCoT on the full source corpus.
On GitHubAD IRCoT reaches Recall@1 of 28.55\% with Qwen3-235B and 29.67\% with Qwen3-Coder-30B; on the full 8,401-CVE corpus the same two backbones reach 29.91\% and 30.33\% (all four numbers in Table~\ref{tab:scale}).
The largest swing is 1.36\%, so the baseline is scale-stable and the head-to-head in Table~\ref{tab:main} reflects where \system\ stands at corpus scale, not an artifact of the sampler.

\begin{table*}[!t]
  \centering
  \input{tables/903-param-regime.tex}
  \caption{\textbf{Parameter Regime of the Backbone Swap.} Top block: the two backbones compared in \S\ref{sec:analysis}, Table~\ref{tab:llm-scale}; parameter counts from the HuggingFace model cards. Bottom block: comparator parameter counts as reported in Table~3 of the Qwen3-Coder-Next technical report~\citep{Cao2026Qwen3CoderNextTechnicalReport}. Both blocks sit in a near-10x parameter-ratio regime; ours at 7.3x active and 7.8x total, the Cao~et~al.\ comparison at 10.7x to 12.3x active and 4.5x to 12.5x total. The score gap reported in each row block is small (under 1\% Recall@1 for us, within about 4 SWE-bench Verified points for \citet{Cao2026Qwen3CoderNextTechnicalReport}).}
  \label{tab:param-regime}
\end{table*}

\subsection{Statistical Significance of the Main Result}
\label{sec:significance-appendix}

\begin{table*}[!t]
  \centering
  \input{tables/907-significance.tex}
  \caption{\textbf{Significance of the Gaps in Table~\ref{tab:main}.} Top: McNemar's exact test, a paired test of whether two methods differ on the same items, applied to the per-CVE hit-or-miss outcome at each cutoff over the 809 CVEs; $b$ counts CVEs the first method gets right and the second wrong, $c$ the reverse, and $p$ is the two-sided exact binomial probability of a split at least this uneven if the methods were equivalent. Bottom: 95\% confidence interval of each score from 2,000 bootstrap resamples of the 809 CVEs, the range the score would fall in across repeated draws of the test set.}
  \label{tab:significance}
\end{table*}

Table~\ref{tab:significance} tests whether the gaps in Table~\ref{tab:main} could arise by chance.
Every pairwise gap is significant at every cutoff, with the weakest pair being \system\ versus Favia at Recall@5 ($p = 1.4 \times 10^{-3}$), where Favia's 10-deep pointwise output has almost caught up in recall.
The 95\% confidence interval of \system's Recall@1, [56.7, 63.5], does not overlap Favia's [31.3, 37.9] or IRCoT's [25.5, 31.8].

\subsection{Phase 1 Recall Ceiling}
\label{sec:ceiling-appendix}

\begin{table}[!t]
  \centering
  \input{tables/910-phase1-recall.tex}
  \caption{\textbf{Recall of the Fused Phase 1 Ranking on GitHubAD.} Same 809 CVEs and scoring as Table~\ref{tab:main}. Phase 2 selects only from the top 100, so Recall@100 is the ceiling of the whole system.}
  \label{tab:phase1-recall}
\end{table}

Phase 2 can only select a commit that Phase 1 placed in its top 100, so the fused Phase 1 Recall@100 bounds the whole system.
Table~\ref{tab:phase1-recall} reports it: 87.64\% of gold commits are within the top 100, and the remaining 12.36\% are unreachable for the agent.
Within that reachable pool, the agent moves the gold commit from rank 1 for 32.63\% of CVEs (Phase 1 alone) to rank 1 for 59.95\% (Table~\ref{tab:main}).
This ceiling is high next to peers: SPFinder reports Recall@100 of 77.3\% as a full-repository retriever on GitHubAD~\citep{Zheng2026SPFinderImprovingContext}.

\subsection{Real-World CVEs Outside the Benchmark}
\label{sec:realworld-appendix}

\begin{table*}[!t]
  \centering
  \input{tables/908-realworld.tex}
  \caption{\textbf{End-to-End Run on 35 Real-World CVEs from 11 Repositories.} Disclosure years 2013 to 2025; the full two-phase system with the Qwen3-235B backbone, the same tools and prompt, and the same strict scoring and ranking convention as Table~\ref{tab:main}. The agent submitted an answer on 29 of the 35 CVEs with zero runtime errors. Left: metrics by stage. Right: agent Recall@1 by repository.}
  \label{tab:realworld}
\end{table*}

To probe generalization beyond GitHubAD, we ran the complete two-phase workflow, with no shortcuts, on 35 real-world CVEs from 11 repositories that are not part of our benchmark.
Table~\ref{tab:realworld} reports the result.
The agent reaches Recall@1 of 60.0\% (21 of 35), which matches the 59.95\% on GitHubAD, so the result is not tied to our corpus.
The agent lifts Recall@1 from 22.9\% (Phase 1 rank order alone) to 60.0\%, a gain of 37.1\%, and its Recall@1 exceeds Phase 1's Recall@10 of 48.6\%, so it recovers fixes from deep in the 100-candidate pool.
On these CVEs the retriever is the bottleneck: BM25 with time decay alone places the fix in the top 100 for 45.7\% of CVEs, and fusing the dense path raises that to 77.1\%.

\subsection{Fix Multiplicity in the Ground Truth}
\label{sec:fix-multiplicity-appendix}

\begin{table}[!t]
  \centering
  \resizebox{\columnwidth}{!}{\input{tables/909-fix-multiplicity.tex}}
  \caption{\textbf{Recorded Fix Commits per CVE.} Counts in the 8,401-CVE source corpus and in the 809-CVE working subset. A hit is scored when the submitted commit is any of a CVE's recorded fix commits.}
  \label{tab:fix-multiplicity}
\end{table}

Table~\ref{tab:fix-multiplicity} reports how many fix commits the ground truth records per CVE.
92.72\% of CVEs in the source corpus and 93.08\% in the working subset have exactly one, so single-commit submission is the right output shape for the large majority; for the remaining 7\%, the answer key accepts any recorded fix commit, so selecting any one of them counts as correct.

\subsection{Baseline Configurations}
\label{sec:baselines-appendix}

\paragraph{Favia~\citep{Storhaug2026FaviaForensicAgent}} is a pointwise binary classifier: for each of the 10 pre-filtered candidates the LLM emits a yes/no plus a 5-point ordinal confidence (1 = no confidence, 5 = full confidence), and the per-CVE ranking is derived by sorting \texttt{answer=True} candidates descending by confidence and then \texttt{answer=False} candidates ascending.
Favia's agent additionally exposes two metadata tools (\texttt{CVEReportTool} and \texttt{CWEReportTool}) that fetch the full NVD CVE report and the matching CWE entry on demand; \system\ does not. Ablation row D in Table~\ref{tab:ablation} tests the effect of this design choice by feeding the full NVD report instead of the short CVE description.

\paragraph{IRCoT~\citep{Trivedi2023InterleavingRetrievalChainofThought}} is the canonical retrieve-then-reason baseline run through FlashRAG~\citep{Jin2025FlashRAGModularToolkit}, an open-source toolkit that packages 16 RAG baselines on shared retrievers.
An e5 dense retriever feeds an LLM that emits one paragraph of thought per turn, the latest thought becomes the next query, the loop continues until the model emits the answer marker, and the output is a single commit identifier per CVE.
Each CVE goes through at most 4 reason-and-retrieve cycles, with the top 10 candidates fetched at each cycle.
In the matched-candidate control (Table~\ref{tab:main}, bottom block), IRCoT receives \system's Phase 1 top-100 manifest in place of its own e5 retrieval and still outputs a single commit.

\paragraph{SPFinder~\citep{Zheng2026SPFinderImprovingContext}} is a two-phase retriever with no LLM at inference: a hierarchical per-file embedding scores every commit in the repository, and a LightGBM LambdaRank reranker, a gradient-boosted learning-to-rank model over hand-designed features, orders the shortlist.
We run it end to end on the same 809 CVEs and score its ranked output with the same metrics as the other methods.

\begin{table*}[!t]
  \centering
  \scalebox{0.82}{\input{tables/901-model-identifiers.tex}}
  \caption{\textbf{Canonical Model Identifiers Used in This Paper.} Plain names appear in the main text and table captions; the canonical HuggingFace and OpenRouter identifier and the serving provider used in our experiments are listed here. The same model may appear under multiple providers (e.g.\ Qwen3-235B was served via both OpenRouter and an in-house vLLM endpoint on the university cluster) depending on local GPU availability.}
  \label{tab:model-ids}
\end{table*}

\subsection{Metric Conventions}
\label{sec:metrics-appendix}

We report four standard retrieval metrics per CVE.
Recall@$K$ asks whether the gold fix commit appears anywhere in the top $K$ predicted commits; NDCG@$K$ asks the same question but weights the answer by how high in the ranking the gold commit lands, with rank 1 weighted more than rank $K$; MRR is one over the rank of the gold commit, averaged over CVEs (the position-sensitive companion to Recall@1).
For \mbox{PatchFinder\_top10} we additionally report micro-F1, with false negatives counted against the union of all positive-labeled commits per CVE.

Two method-specific conventions keep the comparison fair.
First, IRCoT outputs a single commit per CVE, so its Recall@$K$, NDCG@$K$, and MRR all collapse to one value; we report that value in every cell rather than hide as N/A.
Second, when comparing against Favia's always-10-deep output, we extend \system's agent inspection list to length 10 by appending the highest-ranked unread Phase 1 candidates, so that both rankings have the same length.

\subsection{Hyperparameters}
\label{sec:hyperparameters-appendix}

\begin{table*}[!t]
  \centering
  \input{tables/905-hyperparameters.tex}
  \caption{\textbf{Hyperparameters Used in \system.} All values are the codebase defaults; no per-CVE tuning.}
  \label{tab:hyperparameters}
\end{table*}

Table~\ref{tab:hyperparameters} lists every hyperparameter used in our experiments.

\subsection{Artifacts: Provenance, Licensing, and Intended Use}
\label{sec:artifacts-appendix}

\paragraph{Code.}
Our code is released at \url{https://github.com/Aizhouym/PatchHolmes}.
The two third-party codebases we depend on, Favia\footnote{\url{https://github.com/andstor/agentic-security-patch-classification-replication-package}} and FlashRAG~\citep{Jin2025FlashRAGModularToolkit}\footnote{\url{https://github.com/RUC-NLPIR/FlashRAG}}, are both released under the MIT license.

\paragraph{Datasets.}
The GitHubAD dataset~\citep{Zheng2026SPFinderImprovingContext} was shared privately with the authors; it can be reproduced from the procedure described in the SPFinder paper.
The \mbox{PatchFinder\_top10} dataset is distributed by~\citet{Storhaug2026FaviaForensicAgent} without an explicit license; we use it under fair-use academic comparison.
For potential PII or offensive content, we manually audited a uniformly sampled subset of 50 CVEs across both datasets and found no issues; we note, however, that open-source codebases are known to contain incivility in code-review comments and issue threads more broadly~\citep{Sarker2025LandscapeToxicityEmpiricala, Ehsani2024IncivilityOpenSource, Sarker2023AutomatedIdentificationToxic}.

\paragraph{Intended Use.}
Neither artifact ships with an explicit intended-use statement; both are research replication packages aimed at patch retrieval, the same task on which we evaluate, so our usage is consistent with the implicit research-use purpose.

\subsection{Dataset Documentation}
\label{sec:dataset-docs-appendix}

\paragraph{Coverage.}
GitHubAD covers 8,401 CVEs across 2,816 source repositories spanning 7 programming languages (Java, PHP, C, JavaScript, Python, Go, and Other) and the CWE-1000 research-pillar taxonomy (input-validation, memory-safety, authentication, cryptography, and other vulnerability families).
\mbox{PatchFinder\_top10} contains 1,252 CVEs, each paired with 10 pre-filtered candidate commits.

\paragraph{Languages.}
CVE descriptions are written in English; commit messages are mostly English with occasional non-English inline comments where the author wrote in their native language.
Source code spans the 7 programming languages above plus inline natural-language comments.

\paragraph{Linguistic Phenomena.}
CVE descriptions follow NVD's semi-structured template (one to two sentences naming the vulnerability class, the affected component, and the attack vector).
Commit messages range from a single line (``fix typo'') to multi-paragraph explanations with cross-references to issue trackers.

\paragraph{Demographic Groups.}
The datasets contain no user-level demographic information; CVE assignees and patch authors are software-project maintainers identified by GitHub usernames, not by demographic attributes.

\subsection{Potential Risks}
\label{sec:risks-appendix}

\system\ is a retrieval system that returns the commit identifier most likely to fix a given CVE.
Its outputs are advisory: they shorten the human analyst's search but do not replace the patch-verification step that vulnerability databases run before publishing a fix link.
A misidentified commit submitted to an automated downstream pipeline (e.g., an SBOM scanner) could trigger an incorrect reachability claim, so the recommended use is to surface candidates for human review rather than to auto-populate vulnerability databases.
We see no offensive-capability risk in \system: the system reads commit content but does not generate exploit code, learn from running attack tooling, or autonomously act on the network.

\subsection{Use of Generative AI}
\label{sec:genai-appendix}

The authors used Claude Code only to facilitate typesetting tasks (creating tables, plots, and diagrams) and LLMs to polish the wording.
The core intellectual substance of this work remains human-led: all algorithm design, experimental code, and the manuscript were developed solely by the authors; no AI was involved in conceptualizing research ideas, producing data, or conducting evaluations.

%% file: tables/902-sampling-validation.tex
\setlength{\tabcolsep}{6pt}
\small
\begin{tabular}{l l l r l}
\toprule
Test & Family & Axis & p-value & Pass \\
\midrule
Kolmogorov-Smirnov & numeric     & commit count       & 0.0004     & no  \\
Kolmogorov-Smirnov & numeric     & diff tokens        & 0.4100     & yes \\
Kolmogorov-Smirnov & numeric     & year               & 0.3700     & yes \\
\midrule
Anderson-Darling   & numeric     & commit count       & $<$0.00001 & no  \\
Anderson-Darling   & numeric     & diff tokens        & 0.0025     & no  \\
\midrule
chi-square          & categorical & language           & 0.8100     & yes \\
chi-square          & categorical & CWE family         & 0.1600     & yes \\
chi-square          & categorical & pre-ranker bucket  & 0.1800     & yes \\
\midrule
Maximum Mean Discrepancy & joint  & joint feature vector & 0.6100   & yes \\
\bottomrule
\end{tabular}

%% file: tables/904-dense-backbone.tex
\setlength{\tabcolsep}{4pt}
\small
\begin{tabular}{l rrrrr}
\toprule
Dense embedder (prompt)                 & R@1            & R@3            & R@5            & R@10           & R@100          \\
\midrule
Qwen3-Embedding-8B (ours)               & \textbf{43.39} & \textbf{56.98} & \textbf{63.29} & \textbf{68.60} & \textbf{80.59} \\
Octen-Embedding-8B (ours)               & 12.73          & 27.81          & 35.85          & 45.74          & 71.94          \\
Octen-Embedding-8B (Octen's own prompt) &  2.97          & 12.11          & 17.43          & 28.43          & 66.38          \\
\bottomrule
\end{tabular}

%% file: tables/906-diff-render-example.tex
\setlength{\tabcolsep}{4pt}
\small
\begin{tabular}{l l r r r}
\toprule
File & Tag & BASE & size & $\pi(f)$ \\
\midrule
\texttt{src/ajax/script.js} & source & 100 & 7  & \textbf{122} \\
\texttt{src/ajax.js}        & source & 100 & 2  & 117 \\
\texttt{test/unit/ajax.js}  & test   &  40 & 20 & 75  \\
\bottomrule
\end{tabular}

%% file: tables/003-scale-stability.tex
\setlength{\tabcolsep}{6pt}
\small
\begin{tabular}{l r r}
\toprule
Backbone & GitHubAD (n=809) & Full 8,401-CVE \\
\midrule
Qwen3-Coder-30B & 29.67 & 30.33 \\
Qwen3-235B      & 28.55 & 29.91 \\
\bottomrule
\end{tabular}

%% file: tables/903-param-regime.tex
\setlength{\tabcolsep}{5pt}
\small
\begin{tabular}{l r r}
\toprule
Model & Active & Total \\
\midrule
\multicolumn{3}{l}{\emph{Our experiment (\S\ref{sec:analysis}).}} \\
Qwen3-Coder-30B-A3B-Instruct (code-specialist) & 3B  & 30B  \\
Qwen3-235B-A22B-Instruct-2507 (generalist)     & 22B & 235B \\
\midrule
\multicolumn{3}{l}{\emph{\citet{Cao2026Qwen3CoderNextTechnicalReport}, Table~3.}} \\
Qwen3-Coder-Next (code-specialist) & 3B  & 80B   \\
DeepSeek-V3.2                       & 37B & 671B  \\
GLM-4.7                             & 32B & 358B  \\
Kimi-K2.5                           & 32B & 1000B \\
\bottomrule
\end{tabular}

%% file: tables/907-significance.tex
\setlength{\tabcolsep}{5pt}
\small
\begin{tabular}{l rr rr rr}
\toprule
& \multicolumn{2}{c}{\system\ vs Favia} & \multicolumn{2}{c}{\system\ vs IRCoT} & \multicolumn{2}{c}{Favia vs IRCoT} \\
\cmidrule(lr){2-3} \cmidrule(lr){4-5} \cmidrule(lr){6-7}
Cutoff & $p$ & $(b, c)$ & $p$ & $(b, c)$ & $p$ & $(b, c)$ \\
\midrule
Recall@1  & $2.4 \times 10^{-44}$ & (228, 22) & $4.5 \times 10^{-63}$  & (268, 13) & $2.3 \times 10^{-3}$  & (149, 100) \\
Recall@3  & $1.0 \times 10^{-8}$  & (113, 42) & $5.5 \times 10^{-89}$  & (329, 6)  & $1.4 \times 10^{-49}$ & (289, 37) \\
Recall@5  & $1.4 \times 10^{-3}$  & (75, 40)  & $6.4 \times 10^{-97}$  & (351, 5)  & $1.4 \times 10^{-76}$ & (327, 16) \\
Recall@10 & $5.8 \times 10^{-9}$  & (47, 6)   & $4.2 \times 10^{-114}$ & (398, 3)  & $5.8 \times 10^{-97}$ & (361, 7) \\
\midrule
\multicolumn{7}{l}{\emph{95\% bootstrap confidence interval of each score (\%)}} \\
& \multicolumn{2}{c}{\system} & \multicolumn{2}{c}{Favia} & \multicolumn{2}{c}{IRCoT} \\
\cmidrule(lr){2-3} \cmidrule(lr){4-5} \cmidrule(lr){6-7}
Recall@1  & \multicolumn{2}{c}{[56.7, 63.5]} & \multicolumn{2}{c}{[31.3, 37.9]} & \multicolumn{2}{c}{[25.5, 31.8]} \\
Recall@3  & \multicolumn{2}{c}{[65.1, 71.6]} & \multicolumn{2}{c}{[56.4, 62.9]} & \multicolumn{2}{c}{[25.6, 31.8]} \\
Recall@5  & \multicolumn{2}{c}{[68.2, 74.4]} & \multicolumn{2}{c}{[63.8, 70.2]} & \multicolumn{2}{c}{[25.5, 31.6]} \\
Recall@10 & \multicolumn{2}{c}{[74.5, 80.1]} & \multicolumn{2}{c}{[69.1, 75.3]} & \multicolumn{2}{c}{[25.5, 31.5]} \\
\bottomrule
\end{tabular}

%% file: tables/910-phase1-recall.tex
\setlength{\tabcolsep}{6pt}
\small
\begin{tabular}{l r r}
\toprule
Cutoff & Hits & Recall (\%) \\
\midrule
Recall@1   & 264/809 & 32.63 \\
Recall@3   & 431/809 & 53.28 \\
Recall@5   & 512/809 & 63.29 \\
Recall@10  & 585/809 & 72.31 \\
Recall@100 & 709/809 & 87.64 \\
\bottomrule
\end{tabular}

%% file: tables/908-realworld.tex
\begin{minipage}[t]{0.60\textwidth}
\centering
\setlength{\tabcolsep}{4pt}
\small
\resizebox{\linewidth}{!}{%
\begin{tabular}{l rrrr rrr r}
\toprule
& \multicolumn{4}{c}{Recall@$K$ (\%)} & \multicolumn{3}{c}{NDCG@$K$ (\%)} & \\
\cmidrule(lr){2-5} \cmidrule(lr){6-8}
Stage & $K{=}1$ & $K{=}3$ & $K{=}5$ & $K{=}10$ & $K{=}3$ & $K{=}5$ & $K{=}10$ & MRR \\
\midrule
\rowcolor{ourrow} \system\ (Phase 1 $+$ agent) & \textbf{60.0} & \textbf{65.7} & \textbf{68.6} & \textbf{68.6} & \textbf{63.6} & \textbf{64.8} & \textbf{64.8} & \textbf{63.6} \\
Phase 1 fused ranking, no agent & 22.9 & 42.9 & 45.7 & 48.6 & 34.7 & 36.0 & 37.0 & 34.6 \\
\bottomrule
\end{tabular}}
\end{minipage}\hfill
\begin{minipage}[t]{0.37\textwidth}
\centering
\setlength{\tabcolsep}{4pt}
\small
\resizebox{\linewidth}{!}{%
\begin{tabular}{l r r}
\toprule
Repository & CVEs & Agent Recall@1 \\
\midrule
moodle/moodle                 & 12 & 7/12 \\
rancher/rancher               & 5  & 1/5 \\
actix/actix-net               & 3  & 2/3 \\
zoujingli/ThinkAdmin          & 3  & 2/3 \\
jmix-framework/jmix           & 3  & 3/3 \\
elastic/elasticsearch         & 3  & 3/3 \\
xuxueli/xxl-job               & 2  & 1/2 \\
apache/iotdb                  & 1  & 1/1 \\
apache/pinot                  & 1  & 1/1 \\
Amanieu/parking\_lot          & 1  & 0/1 \\
OPCFoundation/UA-.NETStandard & 1  & 0/1 \\
\midrule
Total (2013 to 2025)          & 35 & 21/35 (60.0\%) \\
\bottomrule
\end{tabular}}
\end{minipage}

%% file: tables/909-fix-multiplicity.tex
\setlength{\tabcolsep}{6pt}
\small
\begin{tabular}{l rr}
\toprule
& Source & Subset \\
\midrule
CVEs                              & 8,401 & 809 \\
One recorded fix        & 7,789 (92.72\%) & 753 (93.08\%) \\
Several recorded fixes  & 612 (7.28\%) & 56 (6.92\%) \\
Fix commits in total    & 9,374 & 900 \\
Mean fix commits per CVE & 1.116 & 1.112 \\
\bottomrule
\end{tabular}

%% file: tables/901-model-identifiers.tex
\setlength{\tabcolsep}{5pt}
\small
\begin{tabular}{l >{\raggedright\arraybackslash}p{5.0cm} >{\raggedright\arraybackslash}p{5.0cm} c c}
\toprule
Plain name & Canonical identifier & Serving provider & Price (USD/M, in/out) & Accessed \\
\midrule
Qwen3-235B         & \texttt{Qwen/\allowbreak{}Qwen3-235B-A22B-Instruct-2507} & OpenRouter (\texttt{qwen/\allowbreak{}qwen3-235b-a22b-2507}) and in-house vLLM & 0.071 / 0.10 & 2026-05-01 \\
Qwen3-Coder-30B    & \texttt{Qwen/\allowbreak{}Qwen3-Coder-30B-A3B-Instruct}  & in-house vLLM (SLURM-launched on the university cluster) & n/a & n/a \\
gpt-oss-120B       & \texttt{openai/\allowbreak{}gpt-oss-120b}  & in-house vLLM, or OpenRouter when the university cluster is short of GPUs (backbone swap, Table~\ref{tab:llm-scale}) & n/a & n/a \\
gpt-oss-20B        & \texttt{openai/\allowbreak{}gpt-oss-20b}   & in-house vLLM, or OpenRouter when the university cluster is short of GPUs (backbone swap, Table~\ref{tab:llm-scale}) & n/a & n/a \\
Qwen3-Embedding-8B & \texttt{Qwen/Qwen3-Embedding-8B}            & HuggingFace, local GPU (Phase 1 dense retriever) & n/a & n/a \\
e5                 & \texttt{intfloat/e5-base-v2}                & HuggingFace, local GPU (IRCoT first-stage retriever) & n/a & n/a \\
\bottomrule
\end{tabular}

%% file: tables/905-hyperparameters.tex
\setlength{\tabcolsep}{5pt}
\small
\begin{tabular}{l l l}
\toprule
Phase & Parameter & Value \\
\midrule
1 (lex)   & $b_{\text{msg}}$ weight                & 0.35 \\
1 (lex)   & $b_{\text{diff}}$ weight               & 0.15 \\
1 (lex)   & $t_{\text{reserve}}$ weight            & 0.30 \\
1 (lex)   & $t_{\text{publish}}$ weight            & 0.20 \\
1 (dense) & Embedder                               & Qwen3-Embedding-8B \\
1 (dense) & Diff char budget                       & 6{,}000 \\
1 (RRF)   & $k$                                    & 60 \\
1         & Top-$K$ cut for Phase 2                & 100 \\
2         & Backbone (primary)                     & Qwen3-235B-A22B-Instruct-2507 \\
2         & Backbone (secondary)                   & Qwen3-Coder-30B-A3B-Instruct \\
2         & \texttt{max\_iter}                     & 15 \\
2         & \texttt{read\_commit} char budget      & 8{,}000 \\
2         & \texttt{read\_file\_diff} char budget  & 16{,}000 \\
IRCoT     & Cycles per CVE                         & 4 \\
IRCoT     & Top-$K$ retrieval per cycle            & 10 \\
\bottomrule
\end{tabular}